\documentstyle[epsfig]{aipproc}

\def\gsim{ \lower .75ex \hbox{$\sim$} \llap{\raise .27ex \hbox{$>$}} }
\def\lsim{ \lower .75ex \hbox{$\sim$} \llap{\raise .27ex \hbox{$<$}} }
\def\pp{\noindent\parshape 2 0truecm 16.0truecm 2.0truecm 15truecm}

\begin{document}
\title{The dark matter crisis}

\author{Ben Moore}
\address{Department of Physics, Durham University, UK.}

\maketitle

\


\noindent{\bf Abstract} \  \  
I explore several possible solutions to the ``missing satellites'' 
problem that challenges the collisionless 
cold dark matter model. 


\section*{Introduction}

Most dark matter candidates cannot be distinguished by observations on large
scales.  Although the observed universe appears consistent with hierarchical
structure formation, this still leaves a wide range of potential dark matter
candidates.  For example, the clustering properties of galaxies, abundances of
rich clusters or even halo masses and sizes are all very similar in universes
with matter density dominated by cold dark matter, warm dark matter or
collisional dark matter. We therefore seek tests of the nature of the dark
matter that are sensitive to its interaction properties and small scale power
which manifests itself on non-linear scales.

Cold dark matter (CDM) halos form via a complicated sequence of hierarchical
mergers that lead to a global structure set primarily by violent relaxation.
Numerical simulations have played an important role in determining the shape and
scaling of CDM halo profiles that have subsequently lead to new observational
tests of the model.  A single functional form can fit CDM halos from a mass
scale of $10^7M_\odot$ -- $10^{15}M_\odot$, where the density at a fixed fraction
of the virial radius is higher in lower mass halos and the central profiles
have steep singular cusps (c.f. 
Dubinski \& Carlberg 1991, Warren etal 1992, Navarro etal 1996, Fukushige \& 
Makino 1997, Moore etal 1998, Jing 2000 etc).

Galaxy clusters form via a similar process as individual galaxies, however
most of the galactic fragments that formed clusters have survived the
hierarchical growth, whereas 
on galaxy scales we find little trace of the merging hierarchy. Only a
dozen satellites orbit the Milky Way, whereas a thousand satellites orbit within
the Coma cluster. Numerical 
simulations of CDM halo formation have revealed that the abundance of
dark matter subhalos within a galaxy is the same as found within a scaled galaxy
cluster (Moore etal 1999, Klypin etal 1999). 
There are two solutions to this problem: (i) CDM is incorrect and the
nature of the dark matter suppresses the formation of substructure halos, (ii)
CDM is correct and the dark matter satellites of the Milky Way are present but
only a few percent of them formed stars. 
Here I focus on the latter possibility.

\section{Dark matter substructure}

\centerline{\epsfysize=5.8truein \epsfbox{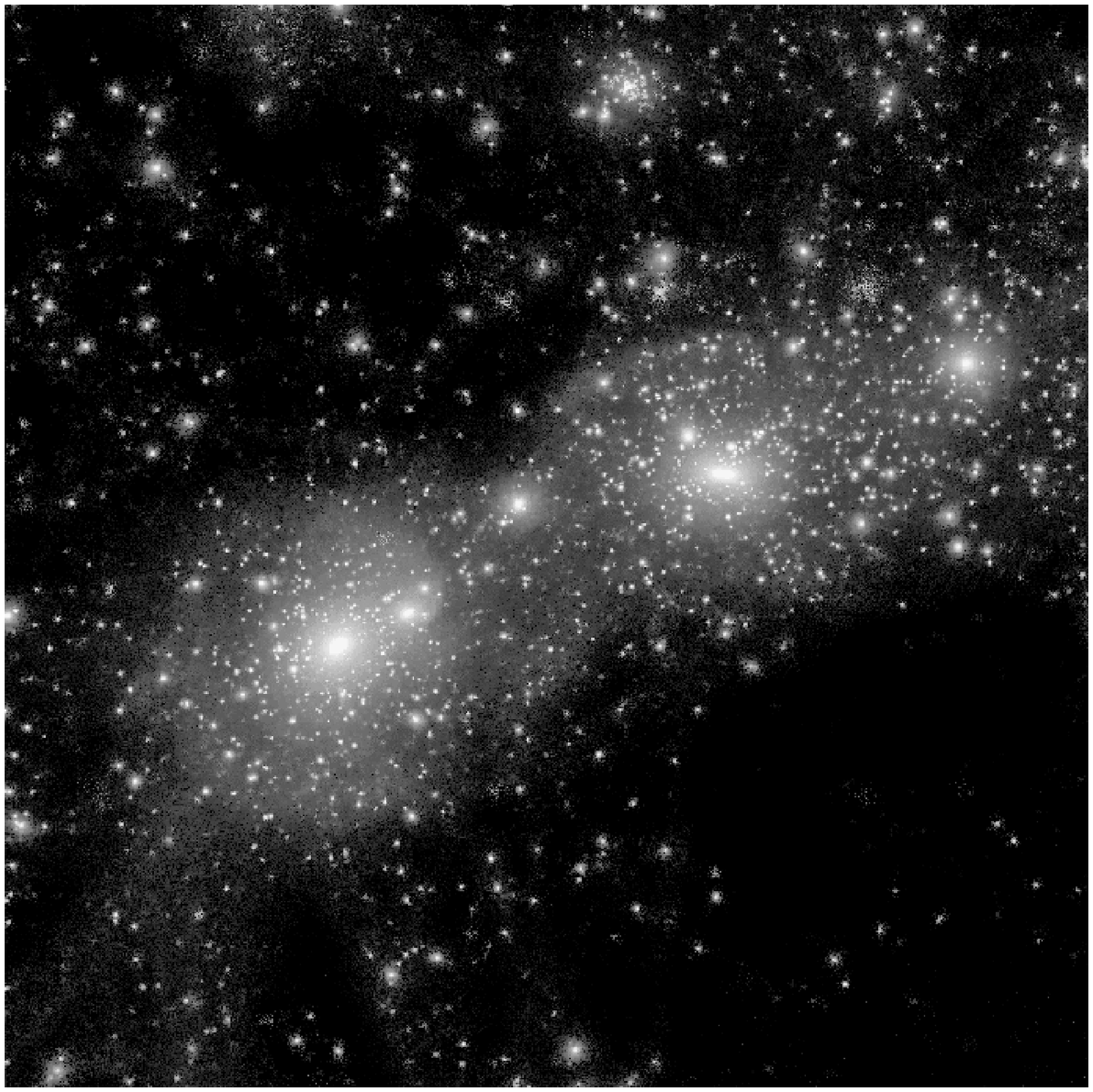}}

\

\noindent{\bf Figure 1.} \ \ The distribution of dark matter with a CDM ``Local
Group''candidate. This is a binary pair of dark matter halos at a redshift z=0,
separated by 1 Mpc and infalling at 100km/s.  The large halos have virial masses of
$\approx 2\times 10^{12}M_\odot$, and 
with a particle mass of $10^6M_\odot$ they
are resolved with over $10^6$ particles and 0.5 kpc force resolution.  The
grey scale represents the local density of dark matter -- there are over 2000
dark matter satellites with circular velocity larger than 10 km/s.

\

\centerline{\epsfysize=4.truein \epsfbox{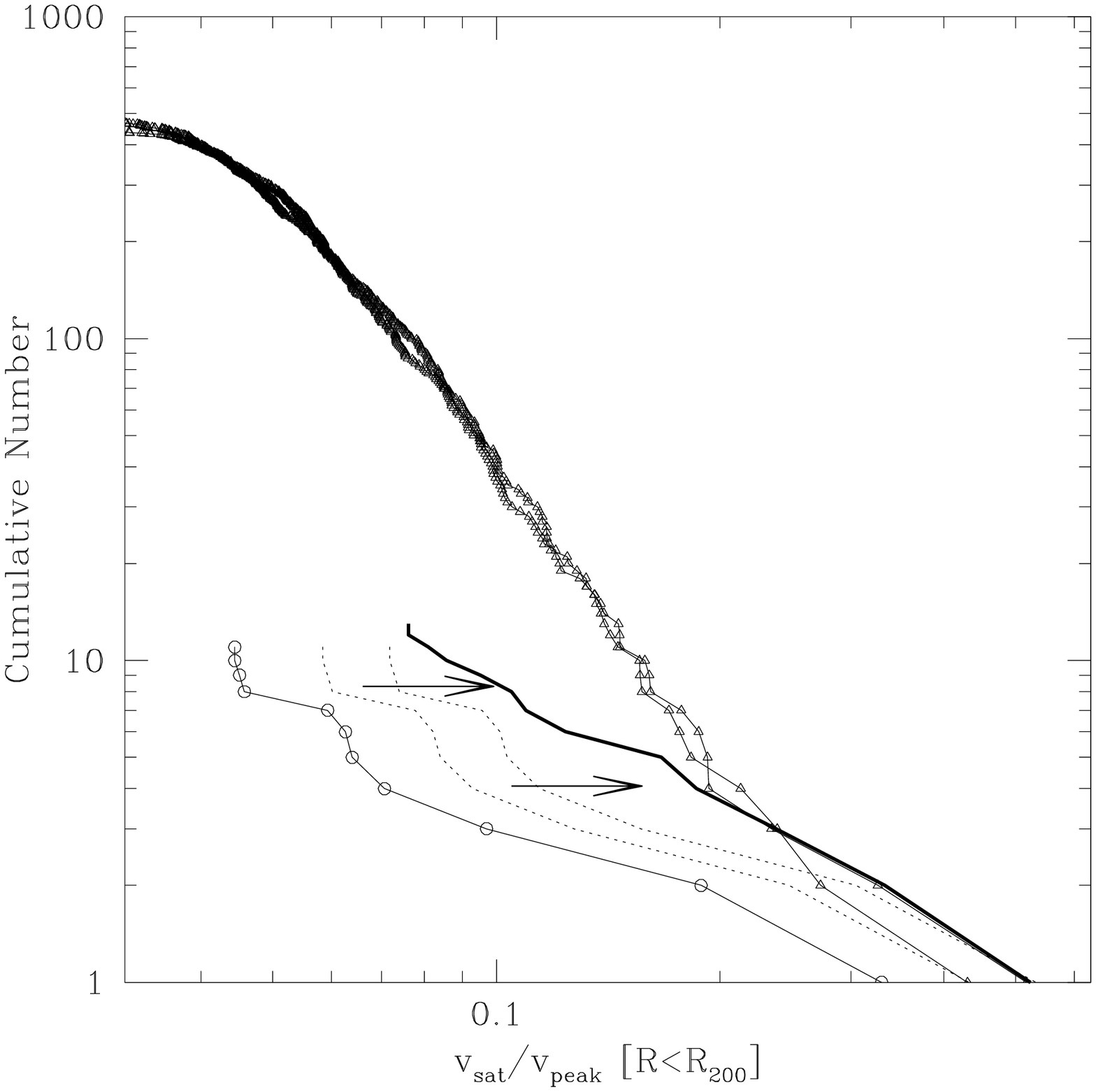}}

\noindent{\bf Figure 2.} \ \ 
The cumulative number of satellites within the virial radius of the Milky Way
(open circles) and within the two CDM halos from Figure 1 (open triangles). Here
I have taken $v_{peak}=210$ km/s for the Milky Way, however the CDM contribution
to the Milky Way can be constrained to lie in the range 130 -- 160 km/s once the
baryonic component has been considered (Moore etal 2001). 
The dotted curves show the effects of
this correction.  The arrows show a correction for converting central velocity
dispersion to $v_{sat}$.  The thick solid
curve shows the distribution of CDM satellites that could form stars before
the universe is re-ionised.

\

\

\noindent{\bf Interpreting the observations}

\

Figure 2 shows the cumulative distribution of subhalos within the high resolution
Local Group halos of Figure 1. The open circles show the observed distribution for
the Milky Way satellites, where I have normalised the distribution 
using $v_{peak}=210$ km/s. However, baryons dominate the central region of the
Galaxy and subtracting the contribution from the disk and bulge gives
the maximum allowed CDM halo that has $v_{peak}=160$ km/s (Moore etal 2001). 
A minimum value of $v_{peak}=130$ km/s is required for the Galactic 
CDM halo to be massive
enough to cool the observed mass of baryons. Figure 2 shows the effect of this
correction.

Simon White has pointed out that the velocity dispersion of the dSph's are
measured well within the cores of their dark matter halos (White 2000). 
We originally
assumed isotropic orbits and isothermal potentials to derive $v_{sat}$ from 
observations of the 1d central velocity dispersion (Moore etal 1999). 
CDM halos have central
density profiles flatter than $r^{-2}$ therefore one expects
the velocity dispersion to drop in the inner region.
M87 provides a good example of this. This galaxy lies at the center of
the Virgo cluster and has a central velocity dispersion of $\approx 350$ km/s
whereas the cluster has a global value that is a factor of two larger.
If we assume that the dSph's are similar to M87, then the correction should
scale roughly as the concentration parameter. Since $c_{M87}/c_{dSph}\approx 
0.4-0.5$,
then we expect the maximum correction to $v_{peak}$ to be an increase of 
50\% over our quoted values. This correction is indicated by
the arrows in Figure 2 which brings the observed data into good agreement with
a crude model for re-ionisation discussed later.




%
%
%
%

\section{Feedback}

\centerline{\epsfysize=4.5truein \epsfbox{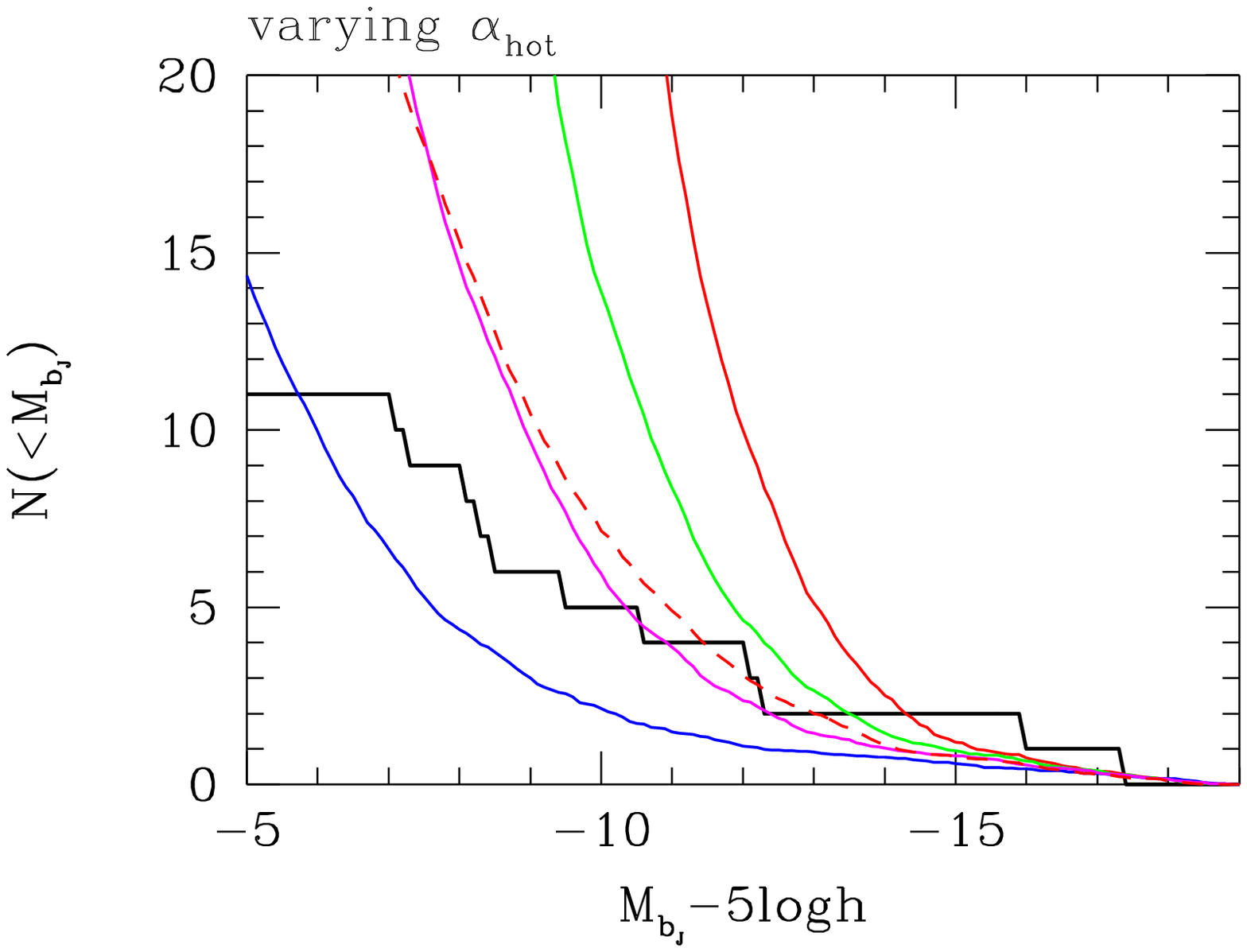}}

\noindent{\bf Figure 3.} \ \
The histogram shows the cumulative distribution of absolute 
magnitudes of the 11 Galactic satellites. The curves show predictions
from the Durham semi-analytic models of galaxy formation (Cole etal 2000) 
varying the parameter $\alpha$ that controls the efficiency of feedback.
This can be tuned to give the correct distribution of satellite luminosities
which in this case would lie somewhere in between the blue and pink curves.

\

Feedback is an essential component of galaxy formation within CDM 
models. It is invoked primarily to flatten the luminosity function given
the steep mass function of CDM halos. For example, the faint 
end of the luminosity function in the Local Group is flat over a 
range that is about 10 magnitudes fainter than $M_*$. By varying the efficiency
of feedback with halo mass, it is possible to get a reasonably flat
luminosity function as Figure 3 demonstrates. The parameter $\alpha$ controls
how much gas is ejected from dark matter halos of a given circular velocity
and allows one to form systematically less stars in smaller mass halos.

The problem with a uniform feedback scheme is that the mass to light
ratios of galaxies will increase rapidly for fainter galaxies.
Thus we find that a satellite halo with absolute magnitude $M_B=-10$ 
is predicted to
have a circular velocity of 40 km/s, roughly three times that observed for
the dSph's. Figure 4 shows the ``Tully-Fisher'' relation for the 11
Galactic satellites compared with the curves predicted from the semi-analytic
models employed in Figure 3. The circular velocities are overestimated
by a factor of 3--4.

\centerline{\epsfysize=4.2truein \epsfbox{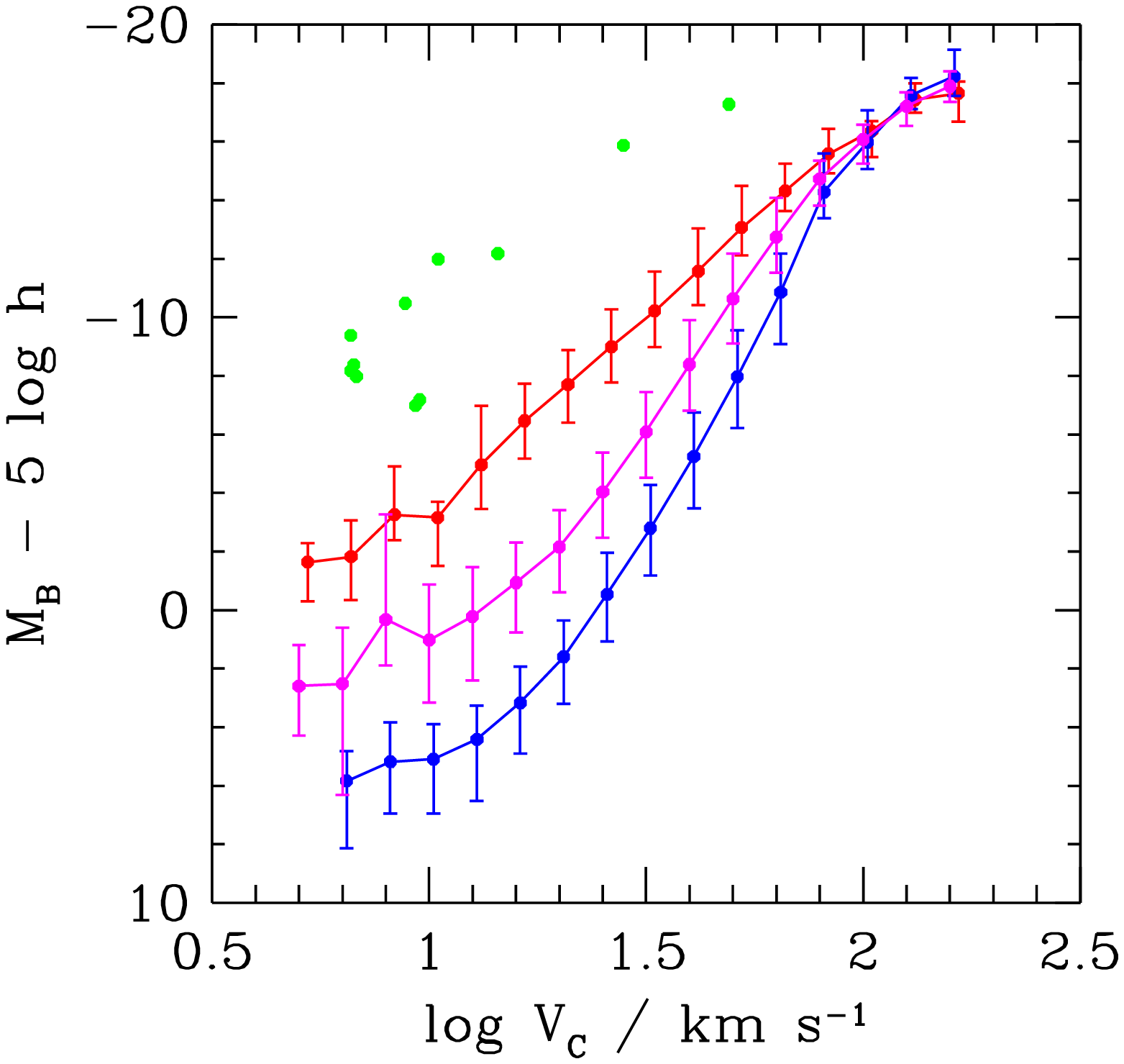}}

\

\noindent{\bf Figure 4.} \ \
The Tully-Fisher relation for the Galactic satellites (green points)
where the velocity dispersions of the spheroidals have been converted
to circular velocity assuming isotropic orbits and isothermal potentials.
The curves show the Tully-Fisher relation predicted by the semi-analytic
models from Figure 3 where we varied the efficiency
of feedback to match the numbers of dwarf galaxies.

\

\section{Re-Ionisation}

\

We have just demonstrated that feedback does not solve the overabundance of CDM
satellites -- clearly some form of stochastic biasing is required.  A solution
was proposed by Bullock etal (2000) in which only those dark matter halos that
have virialised prior to re-ionisation can cool gas and form stars. Once the IGM
has been reheated then the smallest CDM halos cannot capture or cool gas and
they remain completely dark.

In Figure 5 we mark all the progenitor halos that satisfy the condition for
cooling gas prior to z=10, which we will take as the redshift of re-ionisation.
We mark particles red if they lie within a region of overdensity larger than
1000. The locations of these particles are subsequently tracked to z=0 and
marked in the right panel of Figure 5. Roughly 100 satellites satisfy the
density criteria at a redshift z=10 and $\approx 80$ of these physically merger
together to form the very central region of the final galaxy halos.  The
remaining 20 survive intact and can be found orbiting within the virial radius
of the two halos (see Figure 6).  The mean radius of the surviving
satellites is $\approx 80$ kpc, which is a factor of 2.5 smaller than the half
mass radius of the final halos.

\

\centerline{\epsfysize=2.95truein \epsfbox{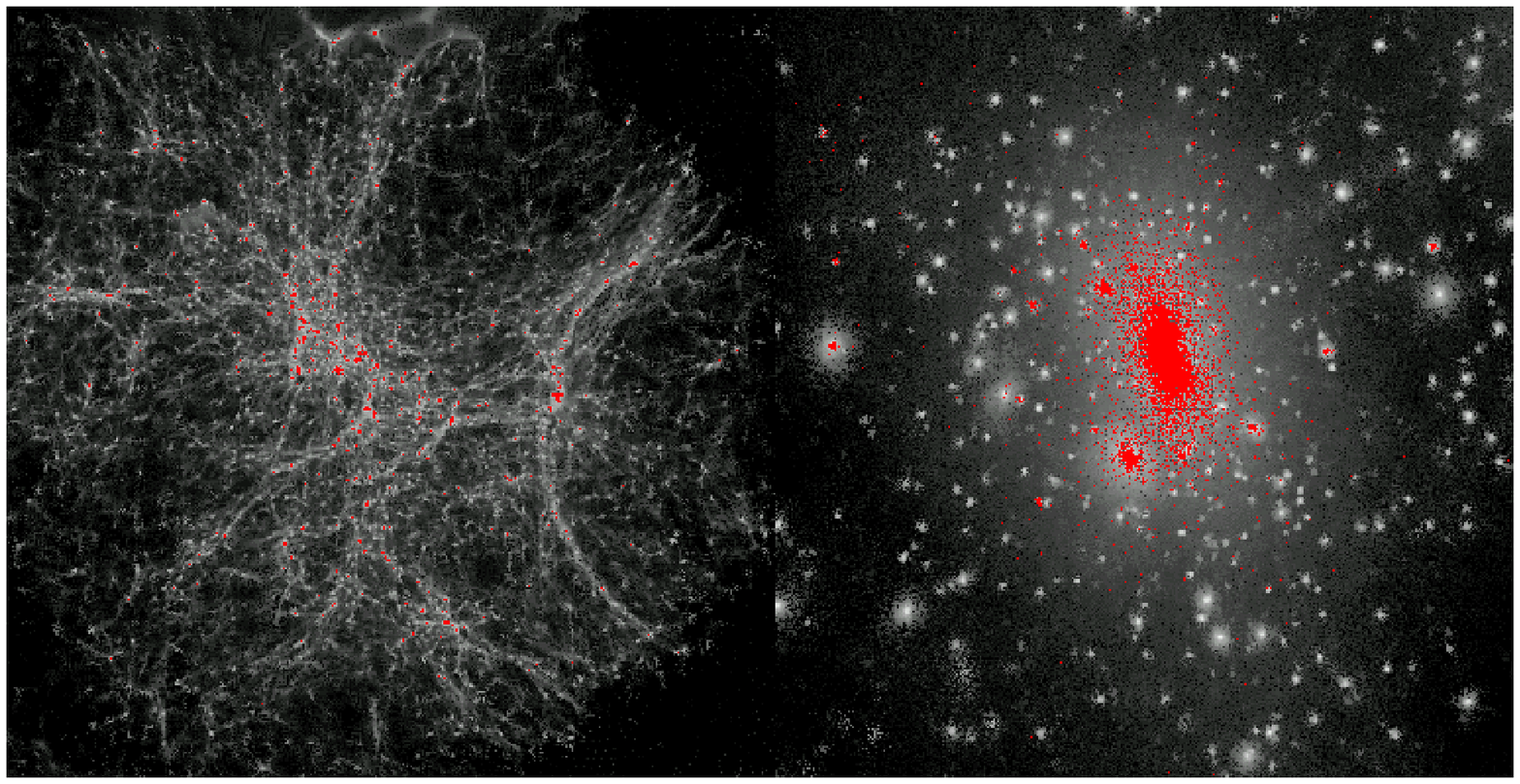}}

\

\noindent{\bf Figure 5.} \ \
The left panel shows the Local Group simulation at z=10. Marked in red
are all those particles that lie in regions with an overdensity 
larger than 1000. The right panel
shows one of the high resolution halos at z=0 and the locations of the
red particles marked at z=10.

\

\centerline{\epsfysize=3.5truein \epsfbox{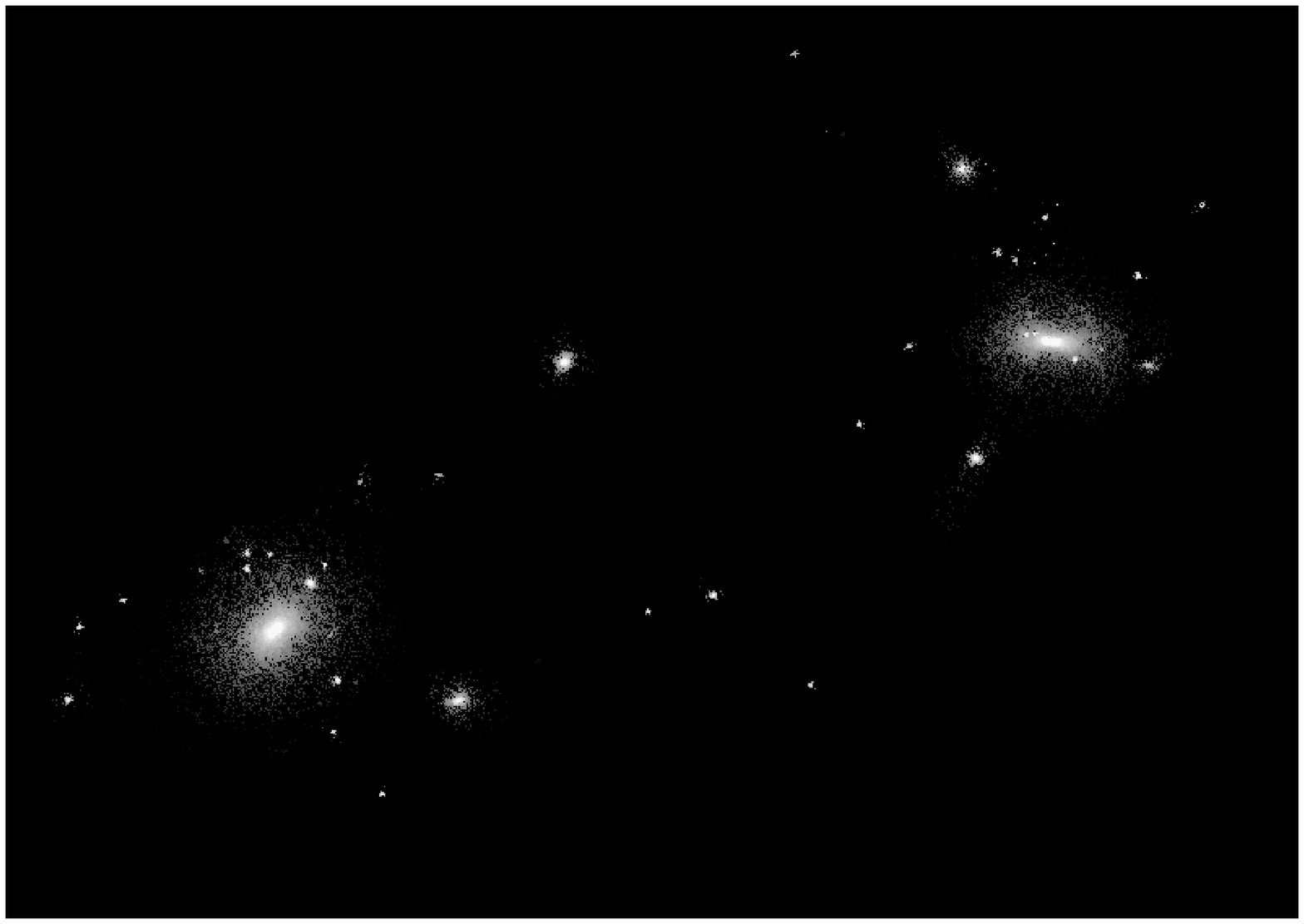}}

\noindent{\bf Figure 6.} \ \
The smoothed distribution of ``starlight'' in the Local Group 
at the present day. I plot only those stars that could form in dark matter 
halos prior to re-ionisation at z=10. The distribution of these stars is highly
biased. Roughly a dozen dark matter dominated satellites orbit within each of
the
parent halos and they have a spatial distribution that matches the real Local Group.
Most of the population II/III stars lie at the very centers of the halos
surrounding M31 and the Galaxy. Their half light radius is just a few
kiloparsecs (c.f. White \& Springel 1999)
and their luminosity density falls as $r^{-3}$ (c.f. Figure 7). 
 
\

The final cumulative distribution of satellites within one of the simulated
halos is shown in Figure 2 and provides a good match to the corrected
observational data points.  Several puzzles remain.  Why don't we find any
satellites in the Galactic halo with velocity dispersion less than $\sim 7$ km/s? Is
cooling that inefficient below $\approx 10$ km/s such that we do not find any
dark matter dominated systems containing just a handful of stars? 

The star
formation histories of the Local Group satellites presents a further
puzzle. Most of the satellites show evidence for several bursts of star
formation, some continuing to the present day. Both re-ionisation and the
``essential'' feedback have been extremely inefficient at removing gas from
these tiny halos that have masses $\approx 10^8M_\odot$.


\vfil\eject

\section{Resolution issues}

{\bf Central halo profiles}

\

Have we really converged on the unique central structure of
CDM halos? This is a hard numerical calculation to
perform since we are always relaxation dominated on small scales.
In a hierarchical universe the first halos to collapse will contain 
just a few particles and have relaxation times much shorter than
a Hubble time.

Figure 7 shows the final density profile of one of the high resolution
Local Group CDM halos. We also plot the final density profile of those particles
that were located within highly non-linear regions at redshifts z=10
and z=20. The central 5 kpc of the halo is dominated by those particles
that were in virialised halos at z=10. Most of these halos
contained just a few particles and their internal structure is
completely dominated by resolution effects. Until we can adequately 
resolve objects collapsing at z=10, we cannot claim to have converged upon
the slope of the density profile at 1--2\% of the virial radius.

\

\centerline{\epsfysize=3.8truein \epsfbox{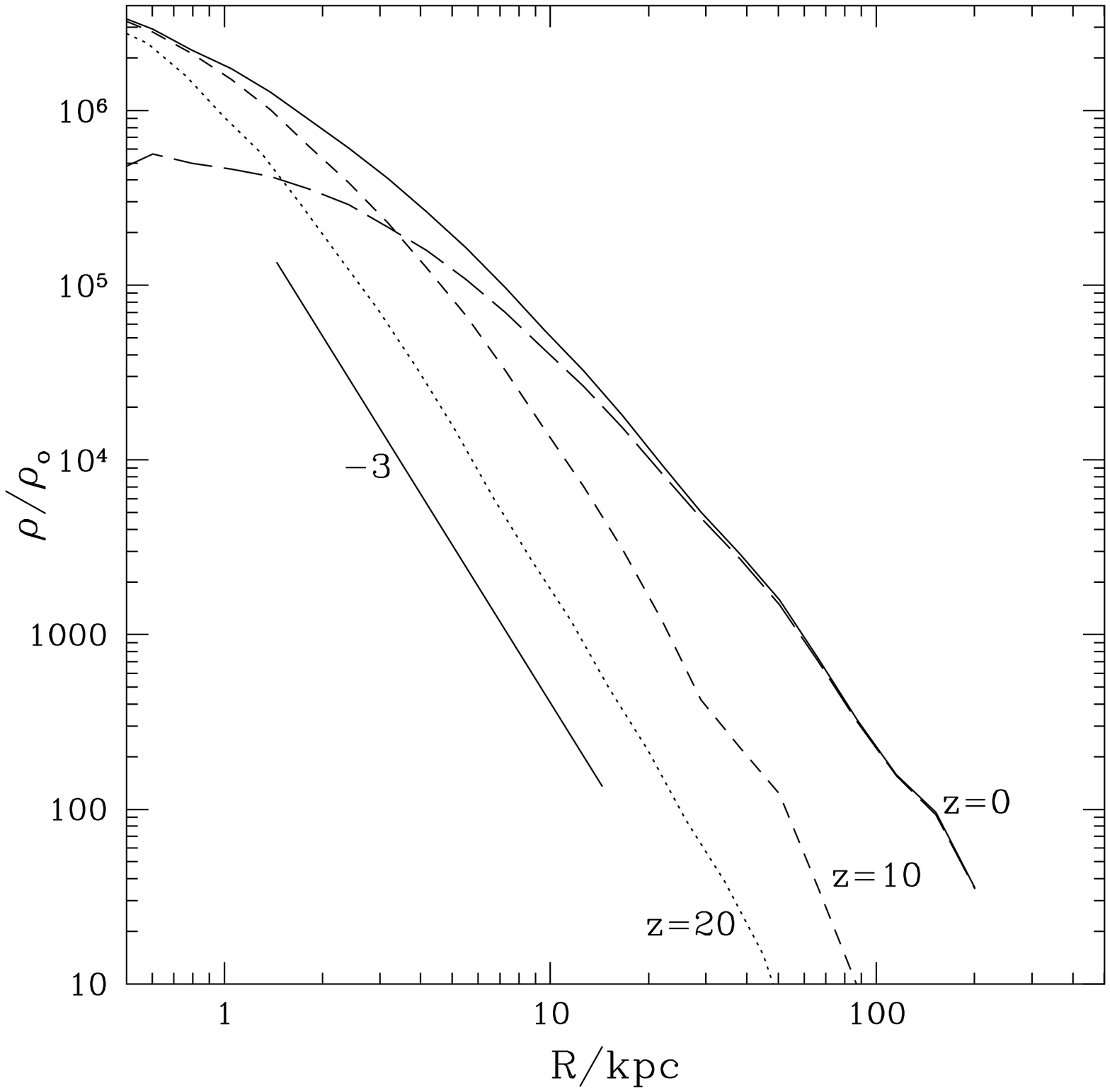}}

\

\noindent{\bf Figure 7.}\ \ 
The solid curve shows the density profile of the high resolution
halo shown in Figure 5. The dotted and dashed curves show the 
density
profiles of those particles that lie in regions of overdensity larger
than 50 at z=20 and z=10 respectively. The long-dashed curve shows the
difference between the solid and short-dashed curves.
The radial density profile of the marked particles at z=0 has a 
gradient of -3 which is similar to that of the Galactic spheroid.

{\bf Beam Smearing}

\

Rotation curves of dwarf galaxies first highlighted potential problems
with the structure of CDM halos (Moore etal 1994, Flores etal 1994, Burkert 1995).
The quality of these data were recently questioned by several authors including
van den Bosch \& Swaters (2000) used rotation curves from 19 dwarf galaxies
to claim that CDM halos are {\it consistent} with the data. However, to make 
this statement these authors had to throw away half of the galaxies and adopt
unphysical (zero) mass to light ratios. Furthermore, seven of the 
remaining nine 
galaxies require concentration parameters in the range c=3--5 which
cannot be obtained in any reasonable $\Lambda CDM$ model.
One could rephrase the conclusions of these authors by stating that
only 2 galaxies from a sample of 19 are consistent with CDM!

Finally, I show the $H_{\alpha}$ and
$HI$ rotation curves of the nearby 
dwarf NGC3109 (Blais-Oullette etal 
2001). These data clearly show that beam
smearing is not an issue for the nearby dwarf galaxies. 
Furthermore, only a constant density
core can fit these data. CDM profiles with central cusps $<-1$  
are ruled out for any value of the concentration parameter.
If CDM is correct then we are forced to conclude that galaxies such as
NGC3109, NGC5585, IC2574, etc, are somehow strange and that their
disk kinematics are somehow not measuring
the mass distribution.

\

\centerline{\epsfysize=3.8truein \epsfbox{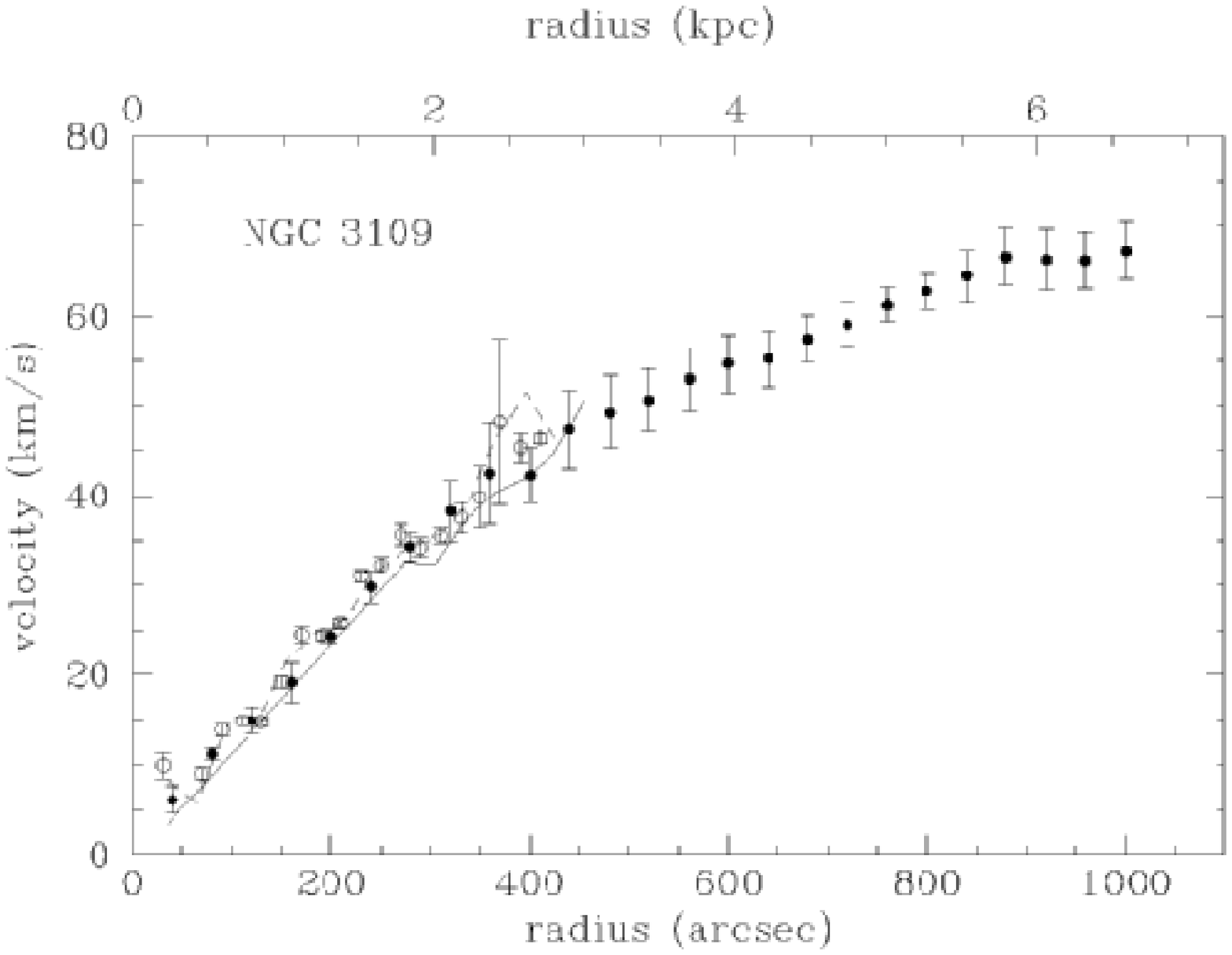}}

\

\noindent{\bf Figure 8.}\ \ The rotation curve of NGC3109 (Blais-Oullette etal 
2000) measured in $HI$ (filled circles) and $H_\alpha$ (open circles). Beam 
smearing is clearly not an issue with nearby dwarf galaxies. 

\

\

\noindent{\bf Acknowledgments} \ \ 
I would like to thank Carlton Baugh for constructing Figure 3 and Figure 4.
BM is supported by the Royal Society.

\

\

\noindent{\bf References}

\

\pp Blais-Ouellette, S., Carignan, C. \& Amram, P. 2000, AJ, in press, astro-ph/0006449.

\pp Bullock, J.S., Kravtsov, A.V. \& Weinberg, D.H. 2000, ApJ, 
astro-ph/0007295.

\pp Bullock, J.S., Kravstov, A.V. \& Weinberg, D.H. 2000, ApJ, 548, 33.

\pp Burkert, A. 1995, Ap.J.Lett., 447, L25.

\pp Cole, S. Lacey, C., Baugh, C. Frenk, C.S. 2000, MNRAS, astro-ph/0007281.

\pp Dubinski, J. \& Carlberg, R. 1991, Ap.J., 378, 496.

\pp Flores, R.A. \& Primack, J.R. 1994, Ap.J.Lett., 457, L5.

\pp Fukushige, T \& Makino, J. 1997, Ap.J.Lett., 477, L9.

\pp Ghigna, S., Moore, B., Governato, F., Lake, G., Quinn, T. \& Stadel, J.
1998, M.N.R.A.S., 300,  146.

\pp Jing, Y.P. 2000, ApJ, 535, 30.

\pp Klypin, A., Kravtsov, A.V., Valenzuela, O., \& Prada, F. 1999,
ApJ, 522, 8.

\pp Moore, B., 1994, Nature, 370, 620.

\pp Moore, B., Ghigna, S., 
Governato, F., Lake, G., Quinn, T., Stadel, J., \& Tozzi, P. 1999, ApJLett, 
524, L19.

\pp Moore, B., Quinn, T., Governato, F., Stadel, J., \& Lake, G. 
1999, MNRAS, 310, 1147.

\pp Moore, B., Calcaneo, C., Quinn, T., Governato, F., Stadel, J., \& Lake, G. 
PhysRevD, submitted.
 
\pp Navarro, J.F., Frenk, C.S. \& White, S.D.M. 1996, Ap.J., 462, 563.

\pp van den Bosch, F.C. \& Swaters, R.A. 2000, AJ, submitted.

\pp Warren, S.W., Quinn, P.J., Salmon, J.K. \& Zurek, H.W. 1992, Ap.J., 
399, 405.

\pp White, S.D.M. 2000, ITP conference, url: online.itp.ucsb.edu/online/galaxy\_c00

\pp White, S.D.M. \& Springel, V. 1999, ESO Workshop, eds A. Weiss, T. Abel \& V. Hill,
astro-ph/9911378.


\end{document}